\documentclass[useAMS,usenatbib]{mn2e}
\usepackage{epsfig}

\begin{document}

\title{On the location of the ice line in circumbinary discs}

\author[M. Shadmehri \& F. Khajenabi ]{Mohsen Shadmehri\thanks{Email:m.shadmehri@gu.ac.ir} and Fazeleh Khajenabi\thanks{E-mail:
f.khajenabi@gu.ac.ir;} \\
Department of Physics, Faculty of Sciences, Golestan University, Gorgan 49138-15739, Iran}

\maketitle

\date{Received ______________ / Accepted _________________ }

\begin{abstract}
Position of the ice line in a circumbinary disc is determined using a simplified and illustrative model. Main sources of the heat in the energy balance of the disc, i.e. heating by the turbulence, irradiation by the components of the binary and the tidal heating are considered. Our goal is to clarify  role of the tidal heating in the position of the ice line. When  viscous heating and irradiation of the binary are considered, ice line lies interior to the inner radius of the disc in  most of the  binaries represented by our parameter survey. But tidal heating significantly extends position of the ice line to a larger radius, so that a smaller fraction of the circumbinaries' population may have ice lines interior to the inner radius of the disc.
\end{abstract}

\section{Introduction}
Despite  a great progress in our understanding of mechanisms of planet formation in the last decade, we are still far from a satisfactory and observationally confirmed theory for the formation of planets. This is  simply because of the complexity and the diversity of the physical processes that are possibly involved. Most of the previous studies in this field have been devoted to a disc around a single star and possible planet formation within it. However, recent discovery of circumbinary planets by Kepler mission  \citep{doyle,oro2,oro1}  is a real challenge to the current theories of the planet formation. Modern attempts are mainly toward extending concepts and methods of the theories of planet formation around single stars to the planet formation in the circumbinary discs \citep*[e.g.,][]{pie,gunter,alex12,rafi13b,rafi13a,martin}.

Although structure of the cicumbinary discs has been studied by several authors using numerical models  \citep*[e.g.,][]{Arty91,Arty94,gunter,pelu,shi}, there are numerical limitations to evolve a circumbinary disc for a long time.  So, attaining a clear physical insight on the structure of a  circumbinary disc can also be provided by  developing semi-analytical models for such a system \citep*[e.g.,][]{Liu,koc,koc2}. Moreover, numerical simulations show a circumbinary  protoplanetary disc resembling to the observed system evolves until it settles in a quasi-equilibrium state \citep{pelu}.  Some authors also  studied time evolution of a circumbinary disc using simplified one-dimensional models to address certain aspects of these systems like photoevaporation \citep*{alex12}, or structure of a layered circumbinary disc \citep*{martin}.

In steady-state semi-analytical models for the structure of circumbinary discs, however, the standard accretion disc model has been extended to include not only the tidal torque on the disc from the central binary but also the tidal heating. \cite*{Liu} constructed a quasistatationary model for a gaseous disc around a binary black hole and showed that the tidal torque creates a gap in the disc near the orbital radius. But they used a simple power-law function of the radius to prescribe viscosity. \cite*{koc} and \cite*{koc2} relaxed this simplifying assumption and their solutions for circumbinary discs are based on the effective viscosity proportional to the gas pressure. However, none of these works included irradiation by the components of the binary.

Temperature distribution in a protoplanetary disc plays an important role in the current theories of planet formation. In particular, it is important to determine locations within a disc with an approximate temperature 160 K where ice can form \citep[see, e.g.,][]{lecar}. Such a region within the disc is called ice line. Interior to the ice line, temperature is so high that icy planets can not form, but  suitable temperature distribution beyond the ice line allows formation of icy planets. Because of the diversity of the physical agents that may affect temperature distribution within a protoplanetary disc, there is considerable uncertainty regarding the exact position of the ice line in a disc. Sources of heat in a protoplanetary disc and possible mechanisms of the energy transport, however, are among the most important physical factors.

Considering  the importance of the ice line in the circumbinary protoplanetray discs and  the effect of the irradiation by the binary, \cite{clanton} presented a simplified but very illustrative model to determine position of the ice line. In this model, only  a detailed energy balance including viscous heating and irradiation from the components of the binary is applied and it was shown that properties of the binary, such as  separation or eccentricity of the system significantly modify position of the ice line. Then, it was suggested that rocky planets should not form in majority of the  binary systems with a total mass less than $4 M_{\odot}$ \citep{clanton}.

Although \cite*{gunter} has already studied irradiated circumbinary discs by doing high-resolution  numerical simulations, neither the position of the ice line nor possible roles of the binary parameters have been investigated. In this regard, \cite{clanton} made a step forward. But an important heating source is the tidal heating which has not been considered in \cite{clanton}. In this paper, we extend model of \cite{clanton} by including the tidal heating in the energy balance of a circumbinary disc and our results show that the position of the ice line is significantly modified in comparison to the  Clanton's model. In the next section,  basic assumptions and general formulation  are presented. Analysis of position of the ice line and its dependence on the binary parameters are discussed in section 3. We conclude by a summary of our results and possible astronomical implications and directions for future work in section 4.

\section{Basic equations and assumptions}
We assume that the disc is in a quasistationary state, where the viscosity and the tidal torques are roughly in balance. Instead of considering all the dynamical equations for a circumbinary disc, just energy equation is analyzed and the rest of the necessary physical quantities such as the surface density and the thickness of the disc are prescribed based on the well-known models. Turbulent heating, irradiation by the binary and the tidal heating are three main sources of the heat which determine the total dissipation rate.  Direct stellar illumination on the exposed inner walls of the truncated outer disc is a form of irradiation from the binary. But in our study stellar irradiation is treated the same as for a continuous disc (as in Chiang \& Goldreich 1997) and that direct stellar illumination, where the angle at which the irradiation strikes the disc does not follow the known relation by Chiang \& Goldreich (1997) (i.e., our equation (5)) is not considered here. Moreover, effect of direct stellar illumination  has already been studied by \cite{Alessio2005} for a disc around a single star. They explained the mid-IR spectrum of a weak line T Tauri star, CoKu Tau/4, based on the emission from the inner wall of a circumstellar disc. Then, it has been discovered that  CoKu Tau/4 is a close binary and its spectrum is explained from the inner wall of a circumbinary disc \citep{nagel}. Although emission from the inner wall of a circumbinary disc has a vital role in temperature distribution, we neglect this type of irradiation to illustrate possible role of the tidal heating in the temperature distribution and the location of the ice line which has not been studied to our knowledge.

 The emission of the disc is approximated as thermal blackbody radiation. Thus, energy equation becomes
\begin{equation}\label{eq:main}
T_{{\rm mid}}^4 = T_{{\rm vis}}^4 + T_{{\rm irr}}^4 + T_{{\rm tid}}^4,
\end{equation}
where $T_{{\rm mid}}$ is the midplane temperature of the circumbinary disc. Also, $T_{{\rm vis}}$, $T_{{\rm irr}}$ and $T_{{\rm tid}}$ are the temperatures due to only viscous dissipation, irradiation flux and the tidal heating, respectively.

\begin{figure}
\epsfig{figure=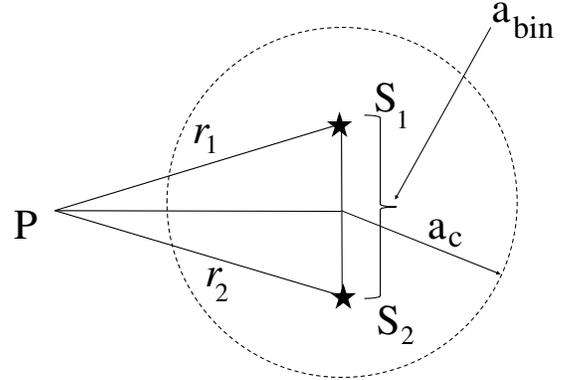,angle=0,scale=0.40}
\caption{Our geometric configuration of the binary is similar to the configuration A in Clanton (2013). A given point P locates at the distances $r_1$ and $r_2$ from the components $S_1$ and $S_2$, respectively. Inner radius of the disc, i.e. $a_{\rm c}$, is shown by a dashed circle. Also, binary separation is denoted by $a_{\rm bin}$.}
\label{fig:f1}
\end{figure}

The following standard equation gives viscous temperature as a function of the radial distance $a$, i.e.
\begin{equation}\label{eq:tem-vis}
T_{{\rm vis}}^4 = \frac{3}{4} (\tau_{\rm R} + \frac{2}{3}) \left (\frac{3}{8\pi \sigma} \frac{GM\dot{M}}{a^3}\right ),
\end{equation}
\citep*[e.g.,][]{bell,clanton}, where masses of the primary and the secondary stars are denoted by $M_1$ and $M_2$ and the total mass is $M=M_1 + M_2$. Also, $\tau_{\rm R}$ is the Rosseland mean opacity, i.e. $\tau_{\rm R} \sim \kappa_{\rm R} \Sigma / 2 $ and the opacity is $\kappa_{\rm R} \simeq 1 {\rm cm}^2 g^{-1}$ and the surface density is $\Sigma = (10^3 {\rm g} {\rm cm}^{-2}) (a/AU)^{-3/2}$ \citep{clanton}.  The accretion rate is assumed to be ${\dot{M}} = 10^{-8} {\rm M_{\odot} yr^{-1}}$. But we also consider lower accretion rates in order to present a parameter study.

Irradiative temperature $T_{\rm irr}$ is determined by the incident flux of each component of the binary. As long as the cooling time-scale of the disc is longer than the binary orbital period, position of the ice line will be static. Motivated by this fact, a  configuration for the relative positions of the stars is proposed by \cite{clanton} to calculate the maximum position of the ice line (see Figure \ref{fig:f1}).   In this configuration, the flux incident at a given point P with the distances $r_1$ and $r_2$ from the primary and the secondary stars becomes
\begin{equation}
D_{\rm irr} = (\frac{\alpha_1}{2}) (\frac{R_1}{r_1})^2 \sigma T_{1}^4 + (\frac{\alpha_2}{2}) (\frac{R_2}{r_2})^2 \sigma T_{2}^4 ,
\end{equation}
and then
\begin{equation}\label{eq:tem-irr}
T_{\rm irr} = (\frac{D_{\rm irr}}{2\sigma})^{1/4},
\end{equation}
where $T_1$ and $T_2$ are the stellar temperatures, and $R_1$ and $R_2$ are the stellar radii. If we denote the semi-major axis of the primary and secondary stars by $a_{1}$ and $a_{2}$, then geometrical calculations give us $r_1 = [a_{1}^2 (1\pm e)^2 + a^2]^{1/2}$ and $r_2 = [a_{2}^2 (1\pm e)^2 + a^2]^{1/2}$ where $e$ is the eccentricity of the binary.  Also, $\alpha_{1}$ and $\alpha_2$ are the angles at which the irradiation strikes the disc \citep*{chiang},
\begin{equation}
\alpha_{1,2} \simeq \frac{0.4 R_{1,2}}{a} + \frac{8}{7} (\frac{T_{1,2}}{8\times 10^6 {\rm K}})^{4/7} (\frac{a}{R_{1,2}})^{2/7}.
\end{equation}
Radius and the temperature of the stars are taken from models of \cite*{sie} at an age of one-tenth the stars' zero-age mean sequence.

Gravitational interaction between the components of the binary and the disc leads to a tidal torque which is approximated per unit mass by
\begin{equation}\label{eq:lambda1}
\Lambda (r, a_{\rm bin}) = \frac{q^2 G M}{2a} (\frac{a_{\rm bin}}{\Delta_{\rm p}})^4,
\end{equation}
\citep{Lin86}, where $\Delta_{\rm p} = {\rm max} (H, |a-a_{\rm bin} |)$, $H$ is the disc scale height and $q=M_2 / M_1$ is the mass ratio of the stars. Also,  $a_{\rm bin}$ is the separation of the binary.  Thus, tidal heating is given by
\begin{equation}
Q_{\rm tid} = (\Omega_{\rm b} - \Omega ) \Lambda \Sigma,
\end{equation}
\citep{good,lodato}, where $\Omega = \sqrt{GM/a^3}$ is the Keplerian angular velocity and $\Omega_{\rm b} = \sqrt{GM/a_{\rm bin}^3}$ is the orbital frequency of the binary. Thus, tidal temperature is written as
\begin{equation}\label{eq:tem-tid}
T_{\rm tid}=(\frac{Q_{\rm tid}}{2\sigma})^{1/4}.
\end{equation}

We note that equation (\ref{eq:lambda1}) is valid as long as the mass ratio $q$ is much smaller than one, i.e. $q\ll 1$. This restriction could be relaxed if we can replace this equation by an analytical relation for the tidal heating in nearly equal mass binaries. Unfortunately, there is not such a relation to our knowledge. Although a few authors applied equation (\ref{eq:lambda1}) to a circumbinary disc with the mass ratio around unity \citep{martin}, we think further numerical or analytical studies should address this issue. None of the previous numerical studies of the circumbinary discs calculate  tidal heating {\it directly } from their simulations. Rather, a torque formula is calculated, the amount of
heating derived from that, and an analytic form is assumed based on
linear perturbation theory.

 We also use equation (\ref{eq:lambda1}) for calculating the binary torque, but we replace the mass ratio $q$ with the ratio of the reduced mass, $M_{1}M_{2}/(M_1 + M_2 )$, to the total mass, $M_1 + M_2 $, which becomes
\begin{equation}\label{qprim}
q' \equiv \frac{M_1 M_2 }{(M_1 +M_2 )^2}.
\end{equation}
It will enable us to consider nearly equal mass  binaries as well. We note that when $M_2\ll M_1 $, then $q'$ tends to $q=M_2 /M_1$. Moreover, in calculating the gravitational potential generated by the binary, the above parameter $q'$ appears \citep[e.g.,][their equation (7)]{fac}. However, we think more detailed analytical studies or numerical calculations are needed to confirm our proposed relation (\ref{qprim}) for calculating tidal heating using equation (\ref{eq:lambda1}).

Upon substituting equations (\ref{eq:tem-vis}), (\ref{eq:tem-irr}) and (\ref{eq:tem-tid}) into equation (\ref{eq:main}), the midplane temperature of the circumbinary disc as a function of the radial distance is obtained. Position of the ice line is defined at a radius where the mid temperature is $T_{\rm mid} \simeq 160 {\rm K}$. Stellar properties and the binary separation and the eccentricity of their orbits are among the most important input parameters. In order to illustrate the  role of the tidal heating, location  of the ice line for a given set of the input parameters is determined for the cases with and without the tidal heating. This analysis is performed in the next section.
\begin{figure}
\epsfig{figure=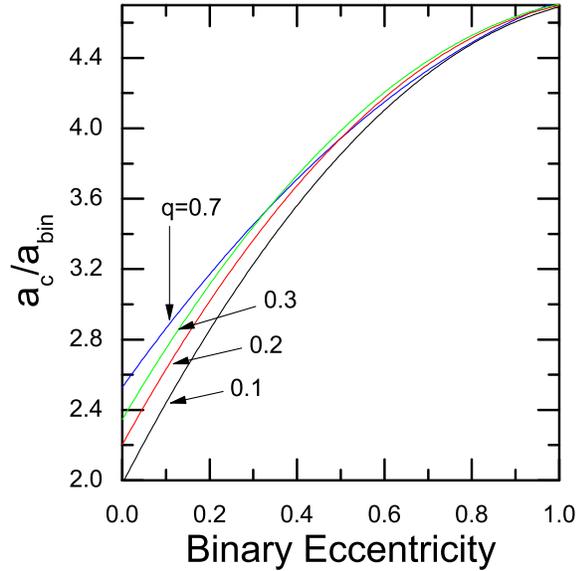,angle=0,scale=0.55}
\caption{Critical semi-major axis $a_{\rm c}$ versus the eccentricity of the binary $e$ based on the equation (\ref{eq:ac}). Each curve is labeled by the mass ratio $q=M_2 / M_1$.}
\label{fig:f2}
\end{figure}

\begin{figure}
\epsfig{figure=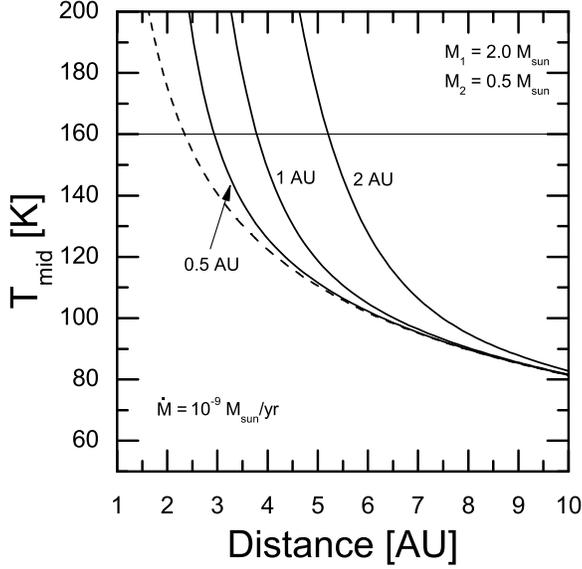,angle=0,scale=0.55}
\caption{Variations of the temperature with the radial distance for a binary system with masses $M_1 = 2  {\rm M}_{\odot}$  $M_2 = 0.5  {\rm M}_{\odot}$ on the circular orbits. The accretion rate is assumed to be $10^{-9} {\rm M_{\odot}/yr}$ and each curve is labeled by the binary separation $a_{\rm bin}$. A case without tidal heating is also shown by dashed line.}
\label{fig:f3}
\end{figure}

\begin{figure}
\epsfig{figure=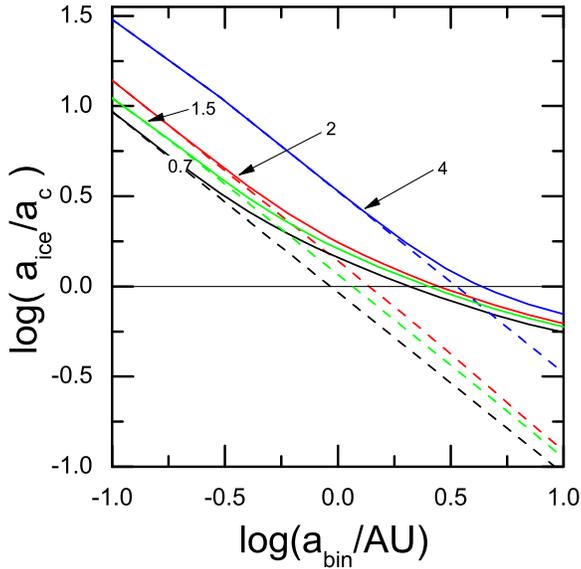,angle=0,scale=0.55}
\caption{Position of ice line normalized by the inner edge radius versus binary separation for unequal mass binaries. Mass of the secondary star is fixed, i.e. $M_2 = 0.5 {\rm M}_{\odot}$. But mass of the primary star increases. Each curve is labeled by the mass of the primary star in solar mass. Dashed curves show position of ice line when tidal heating is neglected and solid curves correspond to the same configuration but including tidal heating. The accretion rate is $10^{-8} {\rm M_{\odot}/yr}$.}
\label{fig:f4}
\end{figure}

\begin{figure}
\epsfig{figure=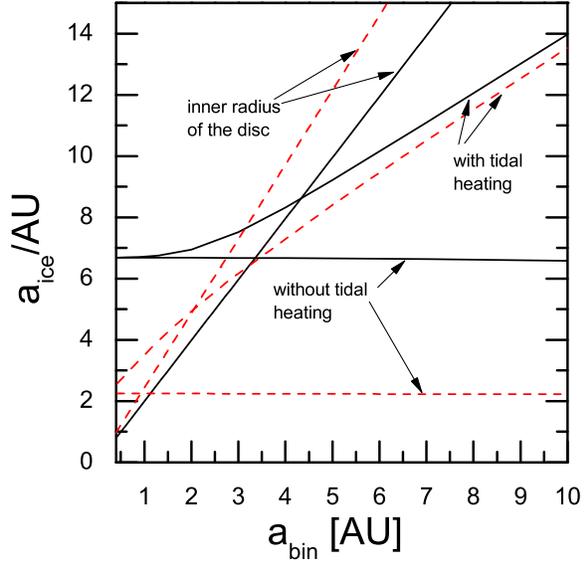,angle=0,scale=0.55}
\caption{Position of ice line versus binary separation for a  binary system with masses $M_1 = 4.0 {\rm M_{\odot}}$ and $M_2 = 0.5 {\rm M_{\odot}}$ (solid line), and another binary system with masses $M_1 = 0.7 {\rm M_{\odot}}$ and $M_2 = 0.5 {\rm M_{\odot}}$ (dashed line). For each case, the inner radius as a function of the binary separation is also shown. The accretion rate is $10^{-8} {\rm M_{\odot}/yr}$.}
\label{fig:f5}
\end{figure}

\begin{figure}
\epsfig{figure=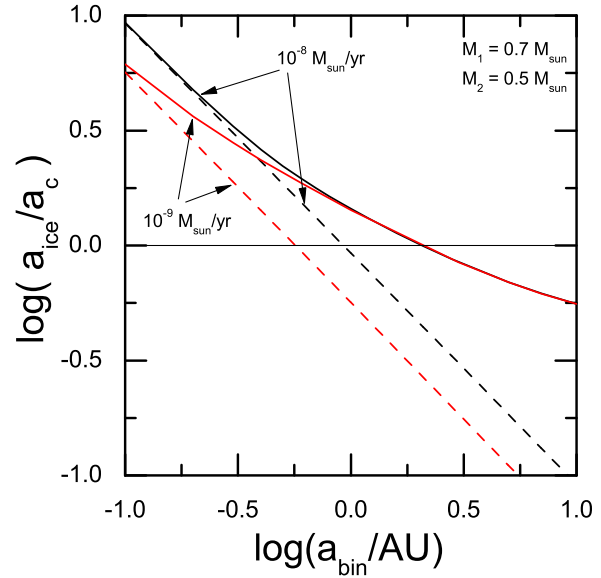,angle=0,scale=0.55}
\caption{ Position of ice line versus binary separation for a  binary system with masses $M_1 = 0.7 {\rm M_{\odot}}$ and $M_2 = 0.5 {\rm M_{\odot}}$, but different accretion rates, i.e. $\dot{M}=10^{-9} {\rm M_{\odot}/yr}$ and $10^{-8} {\rm M_{\odot}/yr}$. Dashed curves show position of ice line when tidal heating is neglected and solid curves correspond to the same configuration but including tidal heating.}
\label{fig:f6}
\end{figure}

\begin{figure}
\epsfig{figure=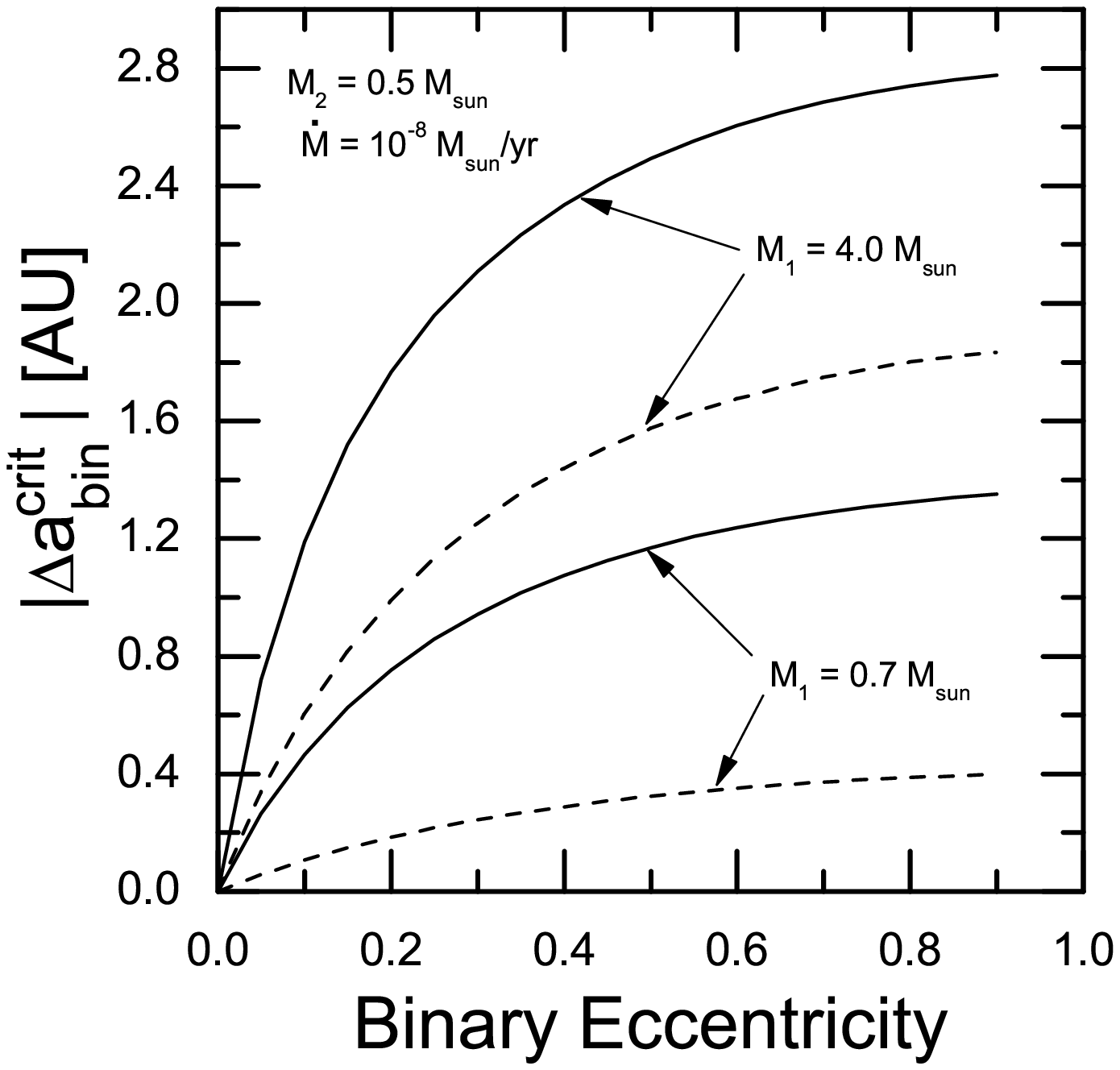,angle=0,scale=0.55}
\caption{Change in the critical binary separation, i.e. $\Delta a_{\rm bin}^{\rm crit} = |a_{\rm bin}^{\rm crit} (e=0) - a_{\rm bin}^{\rm crit} (e)|$, as a function of the binary eccentricity  for unequal mass binaries.  The critical binary separation is defined so that in a binary with a separation larger than $a_{\rm bin}^{\rm crit}$, the ice line is interior to the inner edge of the disc. Mass of the secondary star is fixed, i.e. $M_2 = 0.5 M_{\odot}$. But mass of primary star increases. Each curve is labeled by the mass of primary star in solar mass. Again, solid and dashed curves are corresponding to the cases with and without tidal heating, respectively.}
\label{fig:f7}
\end{figure}

\section{analysis}
We note that the disc is truncated at $a_{\rm c}$ because of the gravitational interaction of the binary with the disc. This inner radius of the disc is a function of mass ratio of the binary and its eccentricity and the separation of the components \citep*[e.g.,][]{Arty94,holman}. For determining  the inner radius of a circumbinary disc, one can consider an ensemble of collisionless test particles in the absence of the gas component \citep*{holman} or in a more physically reasonable approach the fluid nature of gas flow is considered \citep*{Arty94}. \cite{Arty94} showed that the inner edge radius of a circumbinary disc varies from $1.8 a_{\rm bin}$ to $2.6 a_{\rm bin}$ with eccentricity $e$ increasing from 0 to 0.25. On the other hand, \cite{holman} provide the following equation:
\begin{displaymath}
\frac{a_{\rm c}}{a_{\rm bin}} = (1.60\pm 0.04) + (5.01 \pm 0.05) e + (-2.22 \pm 0.11) e^2
\end{displaymath}
\begin{displaymath}
+(4.12 \pm 0.09)\mu + (-4.27 \pm 0.17) e\mu + (-5.09 \pm 0.11) \mu^2
\end{displaymath}
\begin{equation}\label{eq:ac}
+ (4.61 \pm 0.36) e^2 \mu^2 ,
\end{equation}
where $\mu = M_2 /(M_1 + M_2)$.
According to this relation, we can show variations of the inner radius $a_{\rm c}$ versus the eccentricity $e$ for different values of the mass ratio (Figure \ref{fig:f2}).  We emphasize that the above relation for $a_{\rm c}$ is not the same as the disc truncation radius of a gaseous viscous accretion disc. In a gaseous disc the actual truncation radius is closer to the star than $a_{\rm c}$ from equation (\ref{eq:ac}) in many cases because of the effect of viscosity and so, equation (\ref{eq:ac}) provides at best an estimate for the truncation radius of the disc.

Now, we can calculate the temperature distribution and determine location of the ice line. However, as an illustrative example, we present profile of the temperature in Fig. \ref{fig:f3} for unequal mass binaries on the circular orbits. Each curve is labeled by its binary separation. When the tidal heating is neglected, as a reference case, variations of the temperature with the radial distance  is also shown. We see that the temperature drops below a critical value $160 {\rm K}$ at a certain location, i.e. $a_{\rm ice}$.

 Figure \ref{fig:f4} shows the ratio of  the  ice line position $a_{\rm ice}$ and $a_{\rm c}$ as a function of binary separation $a_{\rm bin}$ for unequal mass binaries on the circular orbits. Each curve is labeled by the mass of the primary component in solar mass. Dashed lines correspond to a case without tidal heating and solid lines are for the same case, but including tidal heating.   For a system with a fixed binary separation,  the ice line position shifts toward larger distances from the binary in almost all the considered configurations unless the binary separation becomes too small.  This behavior is explainable if we note that the tidal heating operates as an extra  source of heating. As the mass of the primary star increases, however, curves corresponding to the cases with and without the tidal heating start to deviates from each other at a larger binary separation. For example, when the mass of the primary component of the  binary is four solar masses, the effect of tidal heating on the position of the ice line is almost negligible up to a separation around 1 AU. But in a binary with a less massive primary star, say 1.5 ${\rm M}_{\odot}$, the effect of tidal heating appears at a  smaller separation.

Obviously, position of the ice line can not be less than the inner radius of the disc and this lower limit is shown by a horizontal in Figure \ref{fig:f4}. Each curve intersects this line at a certain separation which is called critical binary separation and is denoted by $a_{\rm bin}^{\rm crit}$. Thus, in a binary with a separation larger than the critical separation $a_{\rm bin}^{\rm crit}$, the ice line is interior to the inner radius of the disc. As Figure \ref{fig:f4} shows tidal heating causes that the critical separation  $a_{\rm bin}^{\rm crit}$ increases in all the considered cases. As we have discussed, there is uncertainty about the exact location of the inner edge of a circumbinary disc. For this reason, we also show variations of  $a_{\rm ice}$ (not normalized by $a_{\rm c}$) versus binary separation in  Fig. \ref{fig:f5} for a binary system with masses $M_1 = 4 {\rm M_{\odot}}$ and $M_{2}=0.5 {\rm M_{\odot}}$ (solid line) and another system with masses $M_1 = 0.7 {\rm M_{\odot}}$ and $M_{2}=0.5 {\rm M_{\odot}}$ (dashed line). Location of the inner edge as a function of the binary separation according to equation (\ref{eq:ac}) has also been shown.

Role of the accretion rate has also been explored in Fig. \ref{fig:f6} where  different accretion rates are considered for the same binary system with masses $M_1 = 0.7 {\rm M_{\odot}}$ and $M_{2}=0.5 {\rm M_{\odot}}$. Solid curves are corresponding to the cases where the tidal heating is accounted for.  Dashed curves correspond  to the same system, but without tidal heating. As the accretion rate decreases, then tidal heating becomes more significant. It is simply because viscous heating is in proportion to the accretion rate, but tidal heating is independent of the accretion rate. Although Figure \ref{fig:f6} is only for a system with fixed masses, we found a similar behavior in the binary systems with different masses.

Effect of binary eccentricity is shown in Figure \ref{fig:f7}. Here, unequal mass binaries are considered. Vertical axis is the difference between the critical binary separation for circular orbits and the critical binary separation for orbits with a given eccentricity, i.e. $\Delta a_{\rm bin}^{\rm crit}=| a_{\rm bin}^{\rm crit}(e=0) - a_{\rm bin}^{\rm crit} (e)|$. Horizontal axis is the eccentricity. Each curve is labeled by the mass of the primary star. Irrespective of the existence of the tidal heating, the parameter $\Delta a_{\rm bin}^{\rm crit}$ increases with the eccentricity of the binary. But variation of $\Delta a_{\rm bin}^{\rm crit}$ is more significant as the mass of the primary component increases. Moreover, the effect of tidal heating in the variation of the critical binary separation is stronger as the orbit of the stars becomes more elongated.

\section{conclusion}
For determining position of the ice line in a circumbinary disc including tidal heating of the binary, a simplified but illustrative model which is a generalization of \cite*{clanton} has been presented. In this study, the energy balance is considered and not only the viscous heating and the irradiation are considered but also it was shown that the tidal heating has a vital role. It is a physical expectation that the tidal heating can extend position of the ice line to the larger radii, but our study provides some quantitative estimates. For  unequal mass binaries, the effect of tidal heating in extending  the location of the ice line becomes more significant as binary separation increases. With decreasing  the mass of the primary star and keeping the mass of the secondary at a fixed value, however, this effect appears even at the  smaller binary separations.

Considering only viscous heating and irradiation, we also found that majority of binaries with a total mass less than four solar masses have ice lines interior to the inner radius of the disc. But for the same explored cases, tidal heating causes ice lines lie at the larger radii beyond the inner radius of the disc. The main suggestion of \cite*{clanton} that rocky planets should not form in these systems   is based on his finding that the position of ice line would be interior to the truncated edge of the disc. In other words, the only planets that can form must be volatile rich and more likely to be ice or gas giants than rocky planet. But this result is modified when tidal heating is considered.  We suggest a smaller fraction of the binaries' population  with the components less massive than two solar masses may have ice lines interior to the inner radius of the disc. So, it is still possible that rocky planets form in these systems.

A more self-consistent analysis, however, is still necessary to confirm our findings. In fact, the energy balance is considered in the absence of the other dynamical equations. For example, surface density is prescribed in our analysis. Gravitational interactions of a binary with its disc not only lead to the tidal heating in the energy balance, but also angular momentum distribution within the disc and consequently other dynamical quantities are modified because of the existence of a tidal torque. Thus, a generalization of the current semi-analytical circumbinary disc solutions \citep{koc2,koc,Liu} to include all three sources of heat is desirable. Such solutions enable us to determine  position of the ice line self-consistently and confirm our results.

\section*{Acknowledgment}
We are grateful to the anonymous referee whose detailed and careful comments helped to improve the quality of this paper.

\bibliographystyle{mn2e}
\bibliography{referenceKH}


%
%
\end{document}